\documentclass[aps,prd,twocolumn,nofootinbib]{revtex4-1}
\usepackage{amsmath,amsfonts,amssymb,graphicx,color,times,bbm,psfrag,dsfont}
\usepackage{amssymb,amsmath}
\usepackage{graphics}
\usepackage{graphicx}

\usepackage[normalem]{ulem}
\usepackage[colorlinks,citecolor=blue]{hyperref}

\def\x{{\bf x}}

\def\pp{{p_\perp}}

\def\qL{{\sqrt{q^2+j \Lambda^2}}}

\newcommand{\be}{\begin{equation}}
\newcommand{\ee}{\end{equation}}

\newcommand{\beq}{\begin{equation}}
\newcommand{\eeq}{\end{equation}}

\newcommand{\Eq}[1]{{Eq.~({\ref{#1}})}}
\newcommand{\Fig}[1]{{Fig.~{\ref{#1}}}}

\begin{document}
\title{Crystalline phases by an improved gradient expansion technique}
\author{S. Carignano}
\affiliation{INFN, Laboratori Nazionali del Gran Sasso, Via G. Acitelli,
22, I-67100 Assergi (AQ), Italy}
\author{F. Anzuini}
\affiliation{INFN and Dipartimento di Fisica, ÒSapienzaÓ Universit\`a di Roma, I-00185 Roma, Italy}
\author{O. Benhar}
\affiliation{INFN and Dipartimento di Fisica, ÒSapienzaÓ Universit\`a di Roma, I-00185 Roma, Italy}
\author{M. Mannarelli}
\affiliation{INFN, Laboratori Nazionali del Gran Sasso, Via G. Acitelli,
22, I-67100 Assergi (AQ), Italy}
\begin{abstract}
We develop an innovative technique for studying inhomogeneous phases with a spontaneous broken symmetry. The method relies on the knowledge of the exact form of the free energy in the homogeneous phase and on a specific gradient expansion of the order parameter. We apply this method to  quark matter at vanishing temperature and large chemical potential, which is expected to be relevant for astrophysical considerations.  The method is remarkably  reliable and  
fast as compared to performing the full numerical diagonalization of the quark Hamiltonian in momentum space, and is designed to improve the standard Ginzburg-Landau expansion close to the phase transition points.  For definiteness we focus on  inhomogeneous chiral symmetry breaking, accurately reproducing known results for 1D and 2D modulations   and examining novel crystalline structures, as well. Consistently with previous results, we find that the energetically favored modulation is the so-called  1D real kink crystal. We propose a qualitative description of the pairing mechanism  to motivate this result.  
\end{abstract}
\maketitle
\section{Introduction}
\label{sec:Introduction}

Quantum Chromodynamics (QCD) at nonzero baryonic densities is expected to exhibit a rich variety of  phases {\cite{Fukushima:2010bq}}.
At sufficiently large values of the quark chemical potential, $\mu$, the chirally broken and confined phase should turn into a chirally restored deconfined  phase. This phase transition is  accompanied by the melting of the chiral condensate and the possible formation of a color superconducting condensate~\cite{Rajagopal:2000}.
These  interesting phenomena are expected to occur in a regime where perturbative QCD computations are insufficient and  ab-initio lattice simulations are currently unavailable due to the sign problem, see for example~\cite{Aarts:2013naa}.  Thus, effective models such as the Nambu--Jona Lasinio (NJL) model are commonly used to describe this region of the phase diagram {(see~\cite{Klevansky:1992qe,Buballa:2005} for reviews)}.

Remarkably,  model calculations indicate that various inhomogeneous phases may arise in quark matter at high density.  Two notable examples are  the crystalline color superconducting phase and  the inhomogeneous chiral  symmetry broken ($\chi$SB) phase  (see~\cite{Anglani:2013gfu,Buballa:2014tba} for reviews). The former is probably located at the onset of the deconfined phase, for neutral and beta equilibrated matter, while the latter 
 is  expected to arise between the
homogeneous $\chi$SB phase and the chirally restored phase if the quark-antiquark coupling strength is sufficiently large.

More specifically, the inhomogeneous $\chi$SB island extends from zero temperature to the chiral critical point, which then turns into a Lifshitz point where three phases (homogeneous $\chi$SB, inhomogeneous $\chi$SB and the chirally restored phase) coexist \cite{Nickel:2009ke}. 
 The onset of this region, separating it from the traditional homogeneous $\chi$SB phase on the left, can be characterized by either a first- or second-order phase transition, depending on the crystalline shape assumed by the chiral condensate~\cite{Buballa:2014tba}. On the other hand, the chiral restoration transition is always found (at least in the chiral limit) to be a second-order one where the chiral condensate gradually melts to zero. A remarkable consequence of this is that the position of the chiral restoration transition is independent from the type of crystalline structure considered. 
  Regarding the color superconducting phases, it is known that at asymptotic densities the {(spatially homogeneous)}
 color-flavor locked phase~\cite{Alford:1998mk} is favored. At densities relevant for compact stars this {homogeneous} phase  could {nevertheless} be superseded by a  crystalline color superconducting phase{, especially when the constraints of charge neutrality and beta equilibrium are considered}. However, whether this phase transition is of the first or second order is not yet established. The transition to the normal phase should be of the second order, although some modulations seem to indicate a first order phase transition. The form of the crystalline pattern has only been semi-quantitatively established in~\cite{Rajagopal:2006ig} by a Ginzburg-Landau (GL) expansion.

Having determined the existence of an inhomogeneous island in the phase diagram, it is natural to ask which crystalline structure will be the most favored one in this region. For definiteness we focus in this work on the  inhomogeneous $\chi$SB phase (but we will comment on applications to the crystalline color superconducting phase, as well). Contrary to the case of crystalline color-superconductivity, mean-field model calculations on the inhomogeneous $\chi$SB phase seem to indicate that the favored  type of modulation is a one-dimensional real structure called a ``real-kink crystal'', which can be expressed in terms of Jacobi elliptic functions~\cite{Thies:2003br,Nickel:2009wj,Buballa:2014tba}. 
{In particular, }
 two-dimensional structures  are found to be strongly disfavored compared to the one-dimensional ones, both in the vicinity of the Lifshitz point~\cite{Abuki:2011pf} 
 as well as at zero temperature~\cite{Carignano:2012sx}. The latter result has been obtained through a computationally expensive analysis. Indeed, even within the NJL model {in the mean-field approximation},  the evaluation of the free energy for inhomogeneous phases is a nontrivial task.
One way to obtain it is by performing a full diagonalization of the quark Hamiltonian {and sum over its eigenvalues}; alternatively GL expansions including gradient terms of the order parameter can be used.
 Both methods, however, have their own limitations: the diagonalization of the Hamiltonian can be performed analytically only in very special cases, while for most types of modulations 
 one has to resort to a brute-force numerical computation in momentum space \cite{Carignano:2012sx,NB:2009}. On the other hand, the GL expansion of the free energy in the order parameter and its gradients is expected to be  valid only close to the second order  transition to the chirally restored phase \cite{Abuki:2011pf,Rajagopal:2006ig,Mannarelli:2006fy}. 
But, gradients of the order parameter may not vanish sufficiently fast even close to the second order phase transition point, making the  GL expansion power counting  nontrivial. One notable exception is the Lifshitz point, where both the order parameter and its gradient are expected to vanish. At any rate, while the determination of the GL coefficients is in principle straightforward, in presence of an inhomogeneous order parameter the actual {derivation of the GL energy functional} becomes extremely tedious at higher orders. The number of possible terms steadily increases and no automated procedure for {the derivation of the coefficients}  has yet been developed. So far, for inhomogeneous chiral condensates  expansions up to the eight order have been derived \cite{Abuki:2011pf}. 

In this work, {with the aim of going beyond these limitations}, we build a controlled framework for investigating inhomogeneous phases away from the Lifshitz point {(for the astrophysical relevant case of matter at vanishing temperature)} without having to resort to a brute-force numerical diagonalization
of the quark Hamiltonian in momentum space. For this, we devise an improved Ginzburg-Landau (IGL) approximation which can correctly describe both phase transitions delimiting the inhomogeneous phase to (at least in principle) arbitrary precision. 
This is done on one hand by implementing ``by construction'' a correct description of the homogeneous phase which also contains information on long wavelength modulations of the chiral condensate, and on the other hand by incorporating a sufficiently large number of appropriate gradient terms. The latter can be determined straightforwardly without having to resort to the full computation of the GL coefficients 
thanks to our analytical knowledge of the eigenvalue spectrum of a simple modulation of the condensate, namely a single plane wave.  

This paper is organized as follows. In Sec.~\ref{sec:IGL} we introduce the IGL approximation to describe
the inhomogeneous phases, focusing on the $\chi$SB case.  In Sec.~\ref{sec:1Dmod} we extract  the coefficients of the IGL expansion governing the transition to the chirally restored phase  {starting from} the single plane wave modulation. In Sec.~\ref{sec:bench} we compare the results of the GL and IGL  approximation with the numerical results {of the diagonalization of the full quark Hamiltonian}.  In Sec.~\ref{sec:twoD} we analyze 2D modulations including a novel ansatz. A qualitative discussion of the pairing mechanism is given in Sec.~\ref{sec:qualitative}. We finally draw our conclusions  in Sec.~\ref{sec:conclusions}.

\section{Improved Ginzburg-Landau expansion} 
\label{sec:IGL}
To develop our formalism we focus on the phenomenon of inhomogeneous $\chi$SB breaking starting from a  GL expansion for the free energy in the NJL model {within the mean-field approximation} \cite{NJL1,Klevansky:1992qe,Buballa:2005}. 

The idea behind the GL expansion is that close to the Lifshitz point the 
thermodynamic potential can be written as an expansion in powers of the chiral order parameter $M(z) = - 2 G [S(z) + i P(z)]$  and its spatial derivatives
 (here $S(z) = \langle \bar\psi \psi \rangle$ and $P(z) = \langle \bar\psi i \gamma^5 \tau^3 \psi\rangle$ are the scalar and pseudoscalar chiral condensates, respectively, and $G$ the scalar coupling constant in the NJL Lagrangian).
More specifically, for a real modulation one can write~\cite{Nickel:2009ke,Abuki:2011pf}

\begin{widetext}
\begin{align}
\label{eq:Omega_GL}
\Omega_\text{GL} = & \Omega[0] + \frac{1}{V} \int d\x\left[   \alpha_2 M^2 + \alpha_4 \left( M^4 + (\nabla M)^2 \right)  + \alpha_6 \left(M^6 + 3(\nabla M)^2 M^2 + \frac{1}{2} ( \nabla M^2 )^2 + 
\frac{1}{2} (\nabla^2 M)^2 \right) \right. \nonumber \\  &\left. +  \alpha_8 \left(M^8+14 M^4 (\nabla M)^2 - \frac{1}{5} (\nabla M)^4 + \frac{18}{5}M (\nabla^2 M) (\nabla M)^2 + \frac{14}{5} M^2 (\nabla^2 M)^2 + \frac{1}{5}  (\nabla^3 M)^2 \right) 
+ \dots \right]\,,
\end{align}
\end{widetext}
where $\alpha_n$ are some coefficients depending on the microscopic model.
The reasoning behind this expansion is that terms with the same $\alpha_n$ are  equally important.
In other words, close to the Lifshitz point both $M$ and $\nabla M$  are small, thus    $M^n$  and $\nabla^m M^{n-m}$, with $n>m$, can be comparable. 

However, this is a very special case, because approaching the second order phase transition $M$ is expected to vanish, but  $ (\nabla M)/M$ may be nonzero. In particular, at $T=0$ and close to the  phase transition to the chirally restored phase the power counting underlying Eq.~\eqref{eq:Omega_GL} is expected to be incorrect \cite{Buballa:2014tba},  more specifically $\nabla^m M^{n-m}$ terms can be larger than the  $M^n$ terms. {In principle}, we expect this to happen  for any sufficiently small temperature away from the Lifshitz point, but in the following we will focus for simplicity on the $T=0$ case, which is relevant for sufficiently old compact stars. Furthermore, \Eq{eq:Omega_GL} is insufficient to provide a reasonable description of the homogeneous $\chi$SB phase.  This calls for a different scheme and/or a different approach. 

From a technical point of view, the GL coefficients $\alpha_n$ in the NJL model can be easily determined, either from their general expression~\cite{Nickel:2009ke},
or more pragmatically by evaluating the free energy for a homogeneous order parameter and performing an expansion in powers of $M$, isolating the coefficients multiplying the $M^n$ terms.
The knowledge of the $\alpha_n$ is, however, insufficient to build a GL functional for inhomogeneous condensates, as different gradient terms of the same order carry different relative prefactors compared to the $M^n$ term (see \Eq{eq:Omega_GL}). {The evaluation of these terms} in the inhomogeneous $\chi$SB phase is already quite tedious at order $\alpha_6$, and becomes increasingly more involved at higher orders~\cite{Abuki:2011pf}.  

To improve the GL scheme and  to circumvent the above technical difficulties, let us take one step back and inspect the structure of~\Eq{eq:Omega_GL}. 
{There we have ordered the terms} in such a way that {order by order} the first one is $M^n$ and the last one  $(\nabla^{\frac{n}2-1} M)^2$.
As already noted, these two sets of terms are particularly relevant: indeed, they are the dominant contributions close to the phase transitions.
In particular,
The  $(\nabla^{\frac{n}2-1} M)^2$ terms are the  dominant gradient contributions  close to the chirally restored phase, because terms with higher power of $M$ are suppressed. 
The  $M^n$ terms on the other hand are particularly relevant close to   to the transition to the homogeneous $\chi$SB phase (indeed these are the only terms present in the homogeneous phase, where gradients vanish).
 However, the free energy of the  homogeneous $\chi$SB  phase is  known in an analytical form.  These consideration motivate the following ``improved Ginzburg-Landau'' (IGL)  expansion, which for a real order parameter reads  \begin{widetext}
\begin{align}
\label{eq:Omega_IGL}
\Omega_\text{IGL}  = &   \frac{1}{V} \int d\x\;  \Bigg[ \, \Omega_\text{hom}(\overline{M^2})  + \alpha_6 \left( 3 (\nabla M)^2 M^2 +
 \frac{1}{2}  (\nabla M^2)^2 \right) \nonumber \\ 
   & + \alpha_8 \left(14 M^4 (\nabla M)^2 - \frac{1}{5} (\nabla M)^4 + \frac{18}{5} M (\nabla^2 M) (\nabla M)^2 + \frac{14}{5} M^2 (\nabla^2 M)^2 \right) 
  + \sum_{n \geq 1} {\tilde\alpha}_{2n+2}  (\nabla^n M)^{2}  \Bigg] \,,
\end{align}
\end{widetext}
where the first and  the last terms in the square bracket  characterize this  novel expansion technique.

{In particular,}  $\Omega_\text{hom}(\overline{M(z)^2})$, is the free energy for an homogeneous order parameter, evaluated  point by point for a moving average of the mass function defined as 
\beq
\overline{M(z)^2}= \frac{1}{\lambda} \int_{z-\lambda/2}^{z+\lambda/2} M^2(\xi) d \xi\,,
\label{eq:mroll}
\eeq
where, as we will see, the relevant wavelength scale $\lambda$ is determined by the chemical potential $\mu$. If $M$ is spatially constant, this term gives by construction the free energy of the homogeneous $\chi$SB phase. {On the other hand, }for a general oscillation it captures the long-wavelength modulation of the condensate: 
 from a point of view of an effective field theory,  this can be seen as the dominant term for long-wavelength fluctuations while high frequency components have been  integrated out. The  $ (\nabla^n M)^{2}$ term is instead the dominant one at high frequencies, of the same order or higher than $\mu$. 
Indeed, the last term in  \Eq{eq:Omega_IGL} includes  the leading gradient contributions close to the second-order transition to the chirally restored phase.  The requirement for this  term to be dominant is the vanishing of the amplitude of the chiral condensate, which is  what happens close to the second order  transition {to the normal phase}.
We labelled the coefficients multiplying these gradient terms as $\tilde\alpha_n$, as they will be equal to the $\alpha_n$ up to a numerical prefactor. By inspecting \Eq{eq:Omega_GL} we can see immediately that $\tilde\alpha_4= \alpha_4$, $\tilde\alpha_6 = \alpha_6/2$ and $\tilde\alpha_8 = \alpha_8/5$, whereas for higher orders these relations have not been determined until now.

The first and the last   term in the  expansion in Eq.~\eqref{eq:Omega_IGL} thus guarantee the agreement with the exact result close to the two phase transitions. 
The other terms, which are taken from the traditional GL expansion, are expected to be relevant in  the  region between the two phase transitions, when   $\nabla M  \sim M^2$, or more precisely when the modulation wavelength, $\lambda$, satisfies $|M| \sim 1/\lambda$. 
As we will see in the following, the $\alpha_6$ terms will be sufficient to provide a good qualitative agreement with the full numerical results, and including the $\alpha_8$ terms will allow us to obtain an excellent quantitative agreement throughout the whole inhomogeneous $\chi$SB phase. 

{The IGL prescription can of course be generalized to complex modulations:
for our novel terms this amounts to simply replacing $M^2$ by $|M|^2$ in
the moving average \Eq{eq:mroll} and  $(\nabla^n M)^2$ by $|\nabla^n M|^2$
in the leading gradient terms.}

One important aspect is that   when derived from an NJL model, the $\alpha_n$ coefficients do not only depend on $\mu$, but  are also sensitive to the regularization scale, $\Lambda$. However,  once this scale is fixed, the coefficients themselves do not depend on the considered {shape of the} modulation {of the  condensate}.  At vanishing temperature
and for a Pauli-Villars regularization with 3 counterterms, a regulator $\Lambda$ and coefficients $c_0=1, c_1=-3, c_2=3,c_3=-1$ (see \cite{Nickel:2009wj})
 we  find   
 \begin{align}
 \label{eq:alpha_n}
 \alpha_2 & = \frac{1}{4G}- \frac{N_f N_c}{8\pi^2}  \left(3 \Lambda^2 \log\left(\frac{4}{3}\right) - 2 \mu^2\right) \,, \nonumber\\
 \alpha_4 & = -\frac{N_f N_c }{16\pi^2} \log\left(\frac{32 \mu^2}{3 \Lambda^2}\right) \,, \nonumber\\
 \alpha_6 & =\frac{N_f N_c }{96\pi^2}\left(\frac{11}{3 \Lambda^2} +  \frac{1}{\mu^2} \right) \,,\nonumber\\
 \alpha_8 & = \frac{ N_f N_c}{256 \pi^2} \left(\frac{1}{2 \mu^4}  - \frac{85}{27 \Lambda^4} \right)  \,, 
\end{align}
where $N_f$ and $N_c$ are the numbers of quark flavors and colors, respectively. 

\section{Higher order gradients from a CDW modulation}
\label{sec:1Dmod}

The only missing ingredients in the IGL  expansion of \Eq{eq:Omega_IGL} are  the $\tilde\alpha_n$ coefficients for $n>8$. We compute them  by exploiting the analytical knowledge of the eigenvalue spectrum of the quark Hamiltonian for  the  so-called chiral density wave (CDW) ansatz
\beq\label{eq:CDW}
M(z) = \Delta e^{2 i q z}\,, 
\eeq
corresponding to a static single plane wave modulation, chosen without loss of generality  along the $z$-direction, with amplitude $\Delta$ and wave number $2q$. 
For this simple case the quasiparticle dispersion law is known \cite{Dautry:1979,NT:2004}:
\beq
E_\epsilon = \sqrt{\pp^2 +\vert E_z+\epsilon q\vert ^2}\,,
\label{eq:ecdw}
\eeq
where $\epsilon=\pm 1$, $\pp=\sqrt{p_x^2+p_y^2}$ and $E_z = \sqrt{p_z^2 + \Delta^2}$. At vanishing temperature the free energy {for this modulation} is given by
\begin{widetext}
\beq\label{eq:Omega_CDW}
\Omega_\text{CDW} = -\frac{N_f N_c }{4\pi^2} \int_0^\infty d\pp \pp \int_{-\infty}^\infty dp_z \sum_{\epsilon=\pm} \Big[ E^\epsilon_{PV} + (\mu - E^{\epsilon})\theta(\mu - E^{\epsilon}) \Big]  +\frac{\Delta^2}{4G}\,,
\eeq
\end{widetext}
where again the Pauli-Villars regularization scheme has been adopted, see \cite{Klevansky:1992qe,Nickel:2009wj,Buballa:2014tba}.  We now expand the free energy  in a Taylor-like series, 
\begin{align}
\Omega &= \Omega_0 + \frac{\partial\Omega}{\partial(\Delta^2)}\Big\vert_{\Delta=0} \Delta^2 + \frac{1}{2}\frac{\partial^2\Omega}{\partial(\Delta^2)^2}\Big\vert_{\Delta=0} \Delta^4 + \dots \nonumber \\ 
&= \Omega_0+ \Omega_2 \Delta^2 + \Omega_4 \Delta^4 +  \dots \,,
\end{align}
starting from $\Omega_0 = \Omega(\Delta=0)$. 
Each term can be separated in a vacuum contribution $\Omega_n^V$ and a medium contribution  $\Omega_n^\mu$ which (at $T=0$) depends on the quark chemical potential $\mu$. For example,  the zero-th order is 
\beq
\Omega_0^\mu = 
  -\frac{N_f N_c }{4\pi^2} \mu^4\,,
\eeq
which is {minus the pressure of a free Fermi gas of massless particles} 
and is $q$-independent, as it should.

The first nontrivial term is  proportional to $\Delta^2$, and is given by  
\begin{widetext}
\begin{align}
\Omega_2 &= \frac{1}{4G} -\frac{N_f N_c  }{4\pi^2}  \lim_{\Delta\to 0}  \int_0^\infty d\pp \pp \int_{-\infty}^\infty dp_z \sum_{\epsilon=\pm} \frac{1}{2 E_\epsilon} \left( 1 + \epsilon \frac{q}{\sqrt{p_z^2 + \Delta^2} } \right) 
\times \left[ \sum_j c_j \frac{E_\epsilon}{\sqrt{E_\epsilon^2 + j \Lambda^2 }}  - \theta(\mu - E_\epsilon) \right] \nonumber\\
 &= \Omega_2^{cond} + \Omega_2^V({0}) +  \Omega_2^V(q) +  \Omega_2^\mu({0}) +  \Omega_2^\mu(q) \,,
 \end{align}
\end{widetext}
where the first term is simply a constant due to the condensation energy, and we explicited the Pauli-Villars regularization of the vacuum part.
Furthermore, we isolated the medium and vacuum $q$-dependent  contributions to $\Omega_2$, which can be evaluated analytically:
\begin{align}
\Omega_2^V(q)  = &-  \frac{N_f N_c }{4\pi^2} q \sum_{j=0}^3 c_j \left[ (q - \qL) \log(j \Lambda^2) \right. \nonumber \\ & \left. + 2 \qL \log(q + \qL) \right]\,,
 \label{eq:o2v}
\end{align}
and
\begin{align}
\Omega_2^\mu(q) = &  \frac{N_f N_c }{4\pi^2}   q \left[ 
 (\mu - q )\log\left(\frac{|\mu - q|}{q}\right) \right. \nonumber \\ &\left. -  (\mu + q) \log\left(\frac{\mu + q}{q}\right) \right]\,.
  \label{eq:o2mu}
 \end{align}
Upon closer inspection, we note that both contributions  carry a  $\log(q)$ term which could possibly spoil any expansion.
 However, by adding them up these log pieces cancel  out. Expanding  in $q/\mu$ we obtain 
\begin{widetext}
\beq
\Omega_2(q) =  \frac{N_f N_c }{4\pi^2}\mu^2 \left[ -\log\left(\frac{32 \mu^2}{3 \Lambda^2}\right) \left(\frac{q}{\mu}\right)^2 + 
\left(\frac{1}{3}+\frac{11 \mu^2}{9 \Lambda^2}  \right) \left(\frac{q}{\mu}\right)^4 + \left(\frac{1}{10} - \frac{17 \mu^4}{27 \Lambda^4} \right) \left(\frac{q}{\mu}\right)^6 +  \left(\frac{1}{21}+\frac{230 \mu^6}{567 \Lambda^6}  \right) \left(\frac{q}{\mu}\right)^8 + \dots  \right]
\label{eq:ome2exp}
\eeq
\end{widetext}
{which is a controlled expansion as long as $q < \mu$. This typically turns out to be the case, as we will see in the following sections. Furthermore, from this expansion } we  can obtain all the $ {\tilde\alpha}_{n}$ terms required for the IGL expansion. Indeed, by inspecting \Eq{eq:Omega_IGL} it is clear that for 
 the CDW all the terms in the form $|\nabla^{\frac{n}{2}} M|^2$ turn into $q^n\Delta^2 $ terms and are therefore all contained in $\Omega_2$. Thus, by expanding $\Omega_2$ in powers
 of $q$ as in \Eq{eq:ome2exp} we can extract these terms to arbitrarily high order in an extremely simple way. 
  Comparing the lower-order coefficients with the expressions in Eq.~\eqref{eq:alpha_n}
we see that they agree, and pushing our expansion to higher orders we find for example
\beq
{\tilde\alpha}_{10} =  \frac{N_f N_c}{1024 \pi^2} \left(\frac{230}{567 \Lambda^6} + \frac{1}{21 \mu^6}\right) \,.
\eeq

 From the above expansion it is clear that the relevant frequencies are of the order of $\mu$,  suggesting  that the scale to be employed in the moving average \Eq{eq:mroll} should be of the order of $1/\mu$.  This scale is also comparable to the radius for the single soliton introduced in \cite{Buballa:2012vm} for the real kink crystal modulation (see later), $R_\text{sol} = \pi/(\sqrt{12}M_\text{vac}) $ ($M_\text{vac}$ being the vacuum constituent quark mass),  since the inhomogeneous phase is typically realized for $\mu \sim M_\text{vac}$.
  In the following we will choose $\lambda = 1/\mu$, although any other choice in the same ballpark leads to similar results.

\section{Benchmarks and applications of the IGL} 
\label{sec:bench}
We are now ready to evaluate and minimize the IGL free energy  and compare it with the standard GL approximation (including up to ${\cal{O}}(\alpha_8)$ terms, see \Eq{eq:Omega_GL}) and the full numerical result. We {begin by considering} two different 1D modulations of the condensate, in order to explore the reliability of the GL and IGL expansions.  
We work in the chiral limit using a Pauli-Villars regulator $\Lambda=757.048$ MeV and a scalar coupling $G=6.002/\Lambda^2$, corresponding to a vacuum constituent quark mass $M_\text{vac}=300$ MeV and a pion decay constant $f_\pi=88$ MeV \cite{Nickel:2009wj}. We remind that all of our results are obtained at zero temperature.

\subsection{Chiral Density Wave}

\begin{figure}
\begin{center}
  \includegraphics[width=.4\textwidth]{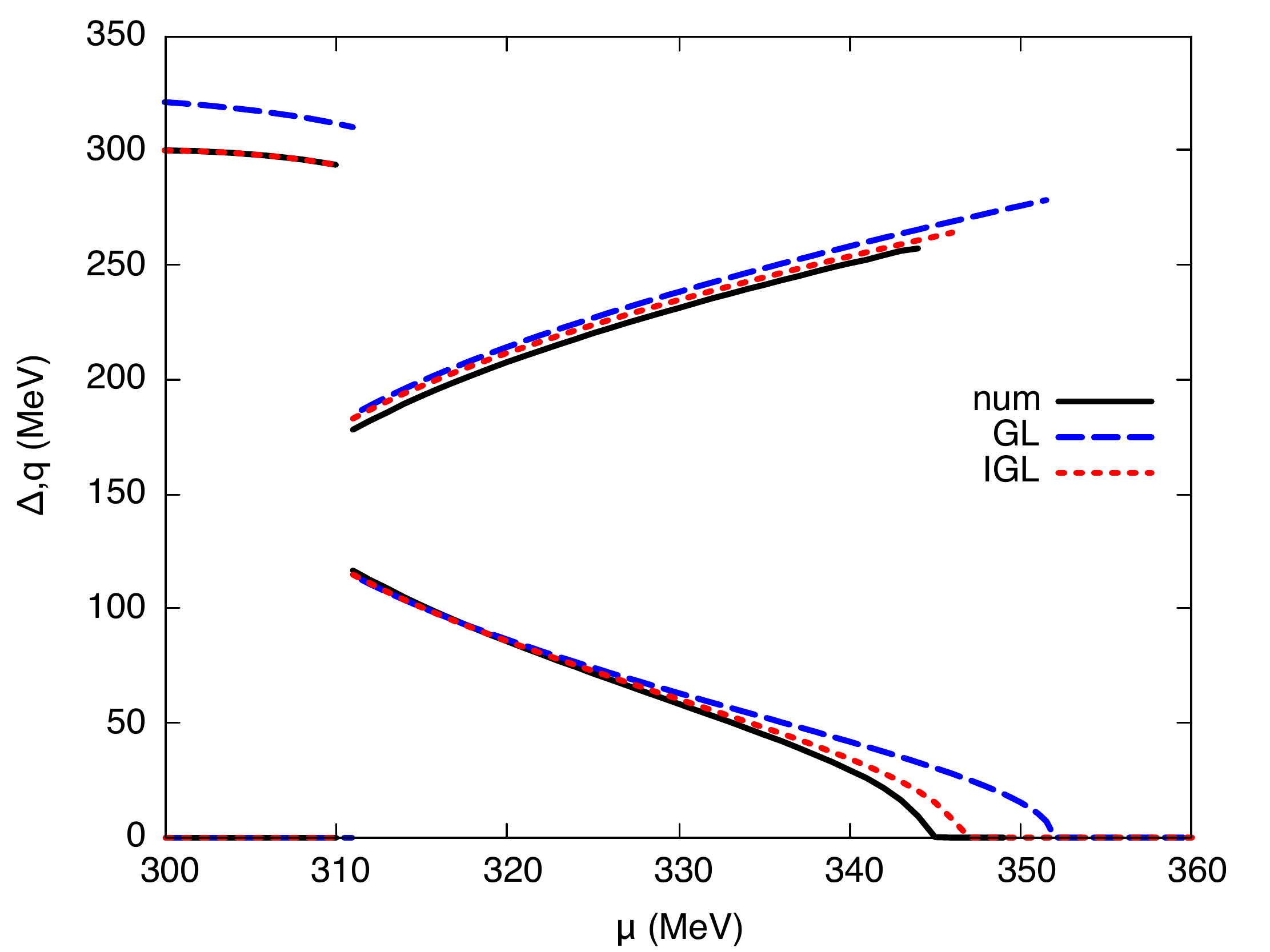}
    \includegraphics[width=.4\textwidth]{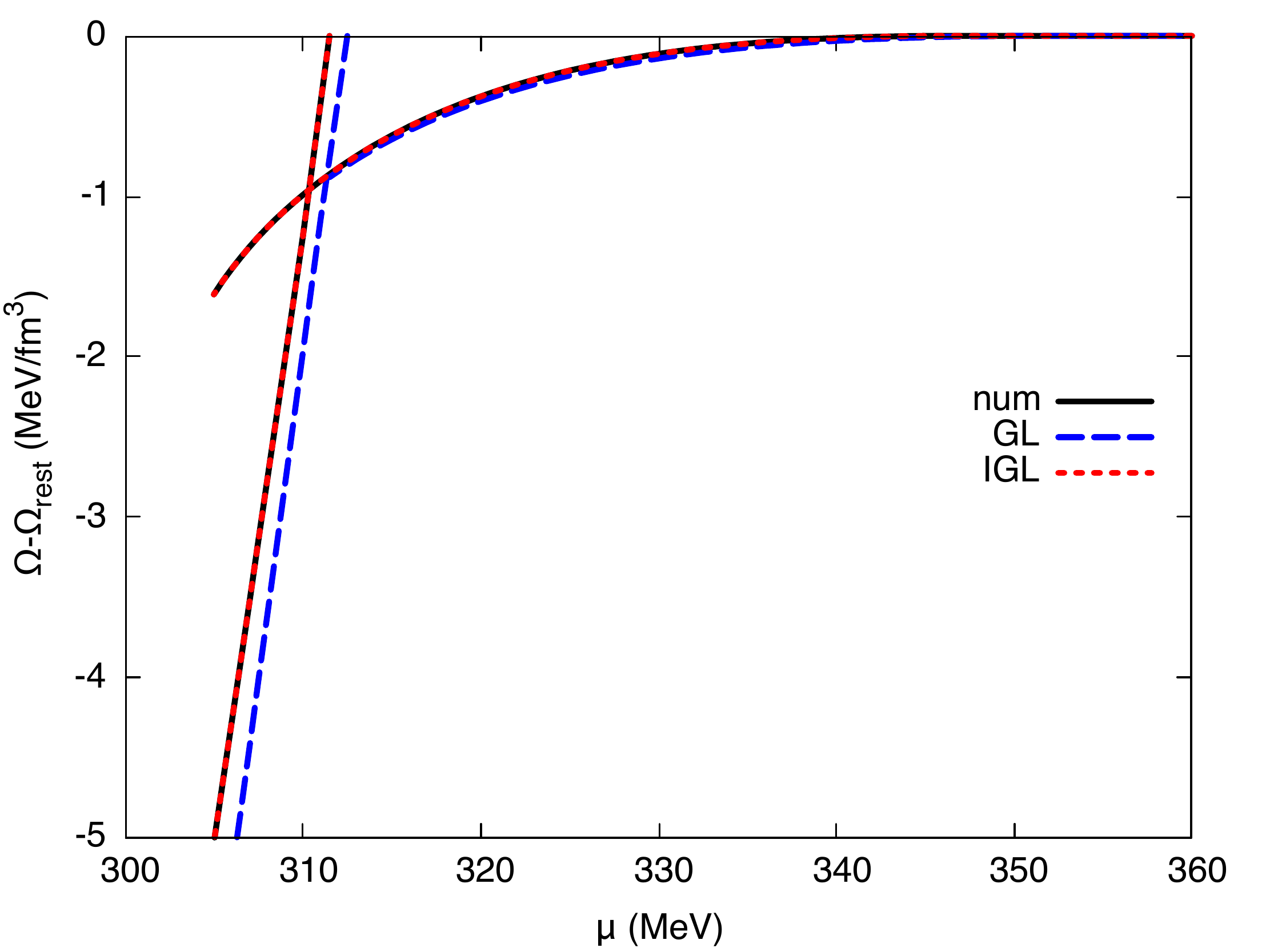}
\caption{Comparison of the numerical results  with  the GL expansion and the IGL approximation  for the CDW modulation as a function of the quark chemical potential.   Top:  values of 
$\Delta$ {(curves with a decreasing behavior)} and $q$ {(curves with an increasing behavior)}
 that minimize the free energy.  Both GL and IGL give good results in the inhomogeneous phase. The GL tends to favor more the inhomogeneous phase over the chirally restored phase and  in the homogenous $\chi$SB phase the GL expansion  tends to overestimate  $\Delta$.   Bottom: free energy {(after subtraction of the free energy of the chirally restored phase, $\Omega_\text{rest}$)}. The approximate expressions almost overlap with the numerical ones in the inhomogeneous $\chi$SB  phase. However, in the homogenous $\chi$SB phase only the IGL approximation leads to a good agreement with the numerical results.
 \label{fig:TFresu1}
}
\end{center}
\end{figure}

We start by considering the CDW modulation, see  Eq.~\eqref{eq:CDW}. In \Fig{fig:TFresu1} we report the results obtained   for the variational parameters $\Delta, q$ (top panel) and the free energy at the minimum (bottom panel).
For the single plane wave, the numerical results (solid black line) are extremely reliable and can be used as a benchmark to test the GL (dashed blue line)  and the IGL (dotted red line)  approximations. {As a first step, we truncated the IGL approximation to order $\tilde\alpha_{10}$.} 
The three approaches give qualitatively similar results, showing that in this case both $\Delta$ and $q$ are discontinuous at the phase transition between the homogeneous and inhomogeneous $\chi$SB phases. The values of $q$ vanishes in the homogeneous $\chi$SB phase and jumps to about $200$ MeV at the onset of the inhomogeneous  $\chi$SB phase. Then it monotonically increases, as a function of $\mu$, in the inhomogeneous phase. The $\Delta$ parameter is instead about $M_\text{vac} = 300$ MeV in the homogenous $\chi$SB phase and then  a decreasing function of $\mu$ in the inhomogeneous phase, eventually vanishing at the second-order  transition to the chirally restored phase.

The first remarkable result visible by inspecting Fig.~\ref{fig:TFresu1} is that even at zero temperature the standard GL expansion provides a good quantitative agreement with the results of the full numerical computation. It fails, however, to  properly reproduce the numerical results in two key regions: Close to the transition to the chirally restored phase,  where it overshoots  the transition point, and at the transition between the  homogeneous and inhomogeneous $\chi$SB  phases,  failing to correctly reproduce the value of $\Delta$  in  the homogeneous $\chi$SB  phase and the transition point.
  On the other hand, the IGL exactly does what it is designed for: it improves the description of these two regions. {Most notably, } it exactly reproduces the free energy in the homogeneous phase and the transition to the inhomogeneous $\chi$SB phase. Moreover, it {shifts the} transition to the chirally restored phase closer  to the numerical result.

\begin{figure}
\begin{center}
\includegraphics[width=.4\textwidth]{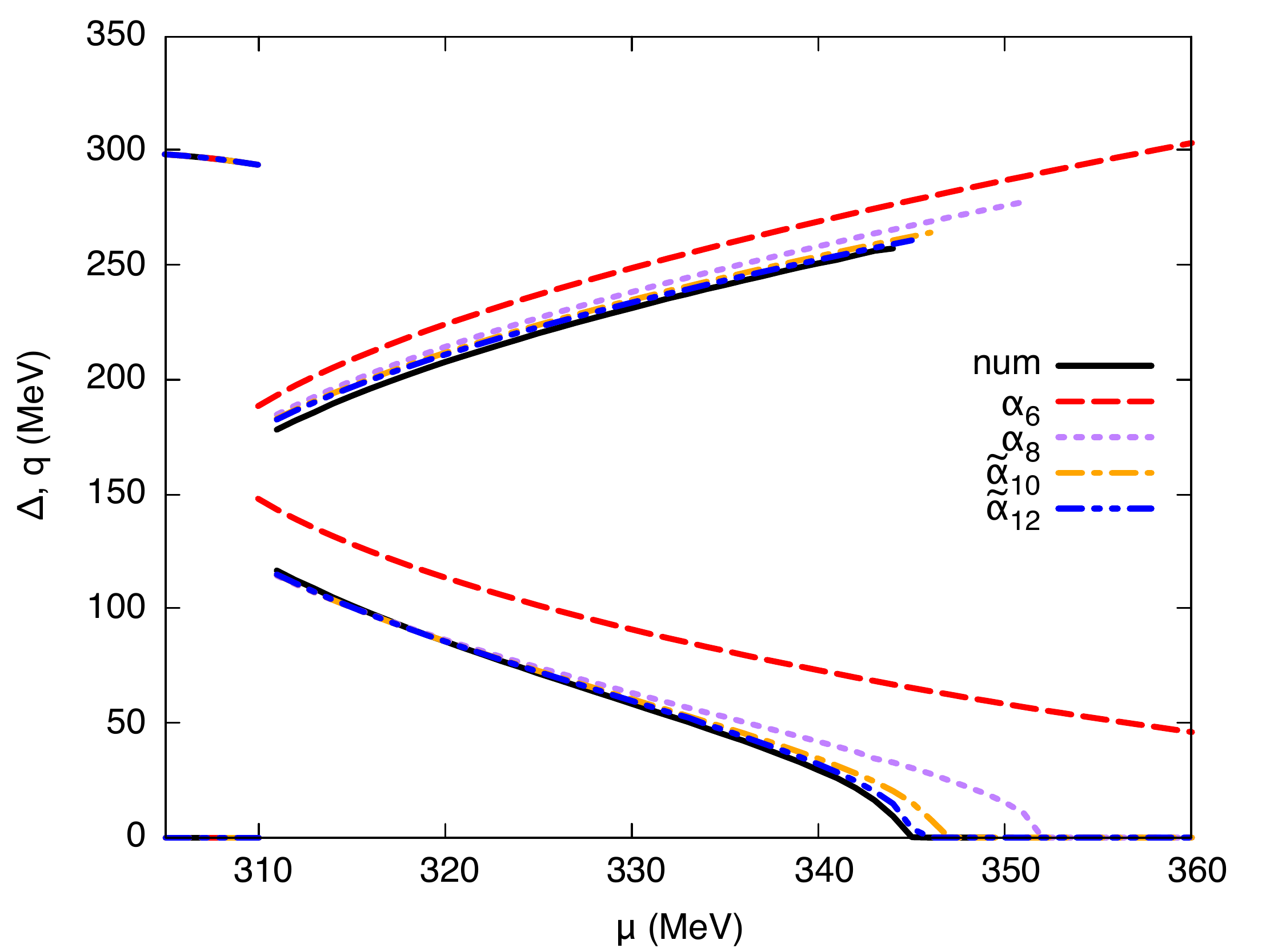}
\caption{Analysis of the IGL approximation for the CDW condensate. The various lines correspond to the values of $\Delta$ (curves with a decreasing behavior) and $q$ (curves with an increasing behavior) that minimize the free energy, as a function of the quark chemical potential. Solid lines are results obtained by the full  diagonalization of the quark Hamiltonian, the others correspond to the IGL expansion, Eq.~\eqref{eq:Omega_IGL}, including  gradient terms of different orders. \label{fig:TFresu2}
 Increasing the number of the  gradient terms, the position of the second order phase transition is increasingly well determined.}
 \end{center}
\end{figure}

It is important to stress that, close to the chiral restoration transition the  IGL is designed for giving a systematic controlled improvement over the standard GL by the inclusion of higher order $|\nabla^n M|^2$ terms, which, as already discussed, can be straightforwardly extracted from \Eq{eq:ome2exp}.
 This is shown in Fig.~\ref{fig:TFresu2}, where we can see how 
the second-order transition is increasingly better described as we include higher order terms. 
In particular, we can see that to get a good qualitative agreement we need at least the ${\cal{O}}(\alpha_8)$ terms, otherwise the inhomogeneous $\chi$SB phase extends to arbitrarily high chemical potentials.  We can interpret this result by inspecting \Eq{eq:ome2exp}, or equivalently the expressions for the GL coefficients: close to the chiral restoration transition and for reasonable values of $\mu/\Lambda$, the leading ${\cal{O}}(\alpha_4)$ coefficient is negative and provides an energy gain in the formation of an inhomogeneous phase, whereas higher order terms constitute energy costs. Indeed, while we find that in principle within our regularization scheme {the coefficients $\alpha_{4n}$ for $n>1$} might flip sign (see  \Eq{eq:alpha_n}) and actually favor large $q$ solutions, in practice this would require chemical potentials too close to the regulator $\Lambda$ for us to trust the model results in that regime\footnote{This behavior might also be related to the appearance of a second inhomogeneous phase at high chemical potentials within NJL model calculations, see the discussions in \cite{Carignano:2011gr,Carignano:2014jla}}.  
Therefore, we find that higher order gradient terms provide (increasingly smaller) energy costs which gradually push the phase transition towards lower chemical potentials, gradually approaching the full numerical result. In practice the convergence of this sum turns out to be very rapid: by including $\tilde\alpha_{10}$ corrections we are off the full numerical result for the transition chemical potential by only 2 MeV, and the IGL results from order $\tilde\alpha_{12}$ on become practically indistinguishable from the numerical result.    
{In light of this, in the following we will consider for simplicity the truncated IGL expansion at order $\tilde\alpha_{10}$, with the understanding that  more refined results can be straightforwardly obtained by simply adding higher order gradient terms.}

\subsection{Real Kink Crystal}
\label{sec:soliton}
The results with the CDW ansatz suggest that the IGL approximation works extremely well. As a second check we compare  the GL and IGL results with the numerical ones for 
the modulation which has been found to be the most favored in the inhomogeneous $\chi$SB window, namely the real kink crystal (RKC) \cite{Nickel:2009wj,Buballa:2014tba}
\be
M(z) = \Delta \sqrt{\nu}\, {\text {sn}}(\Delta z | \nu)\,,
\label{eq:rkc1d}
\ee 
where ${\text {sn}}(\Delta z | \nu)$ is a Jacobi elliptic function, whose shape is characterized by $\Delta$ and by the so-called elliptic modulus $\nu$.

\begin{figure}
\begin{center}
  \includegraphics[width=.4\textwidth]{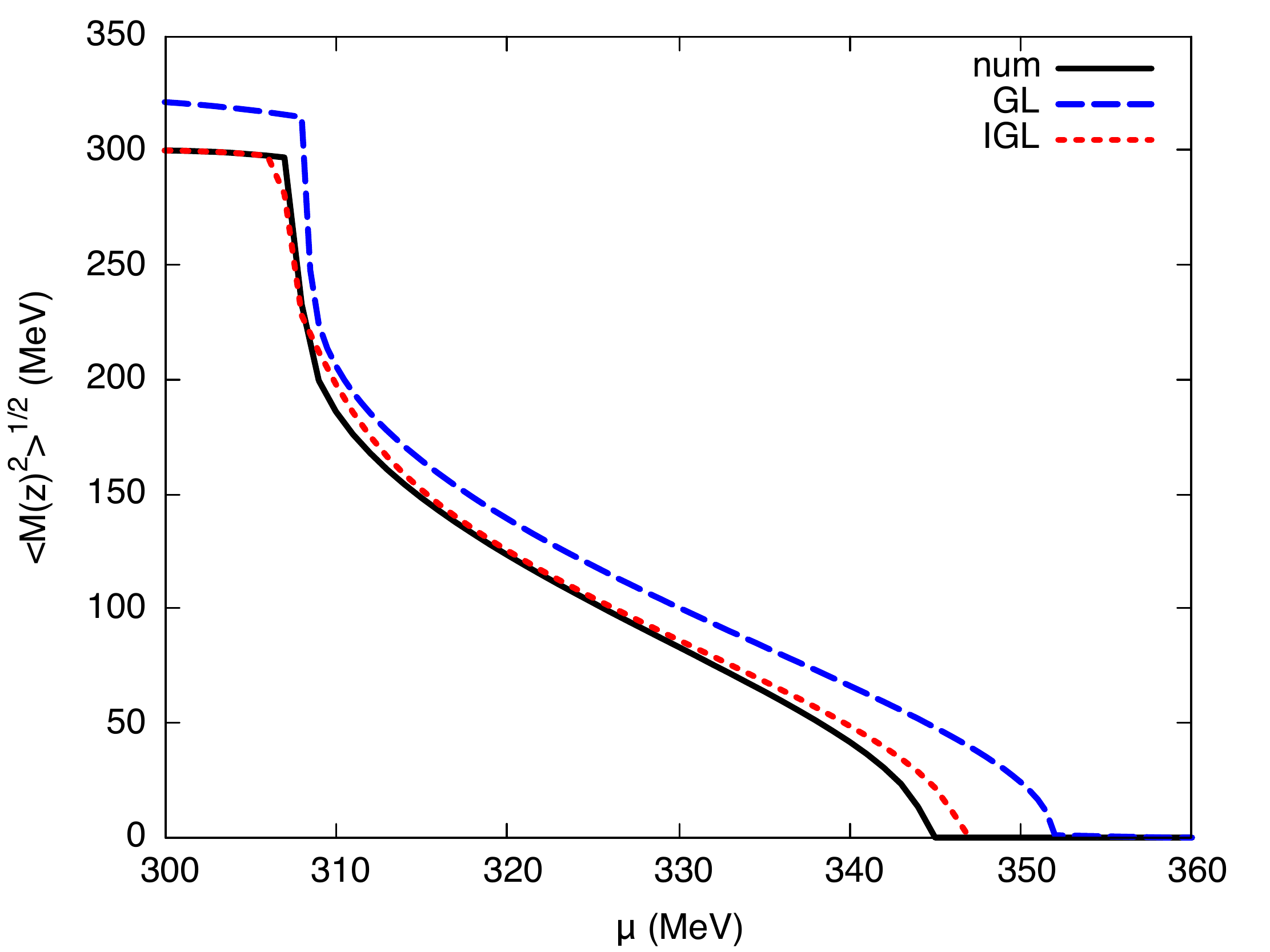}
    \includegraphics[width=.4\textwidth]{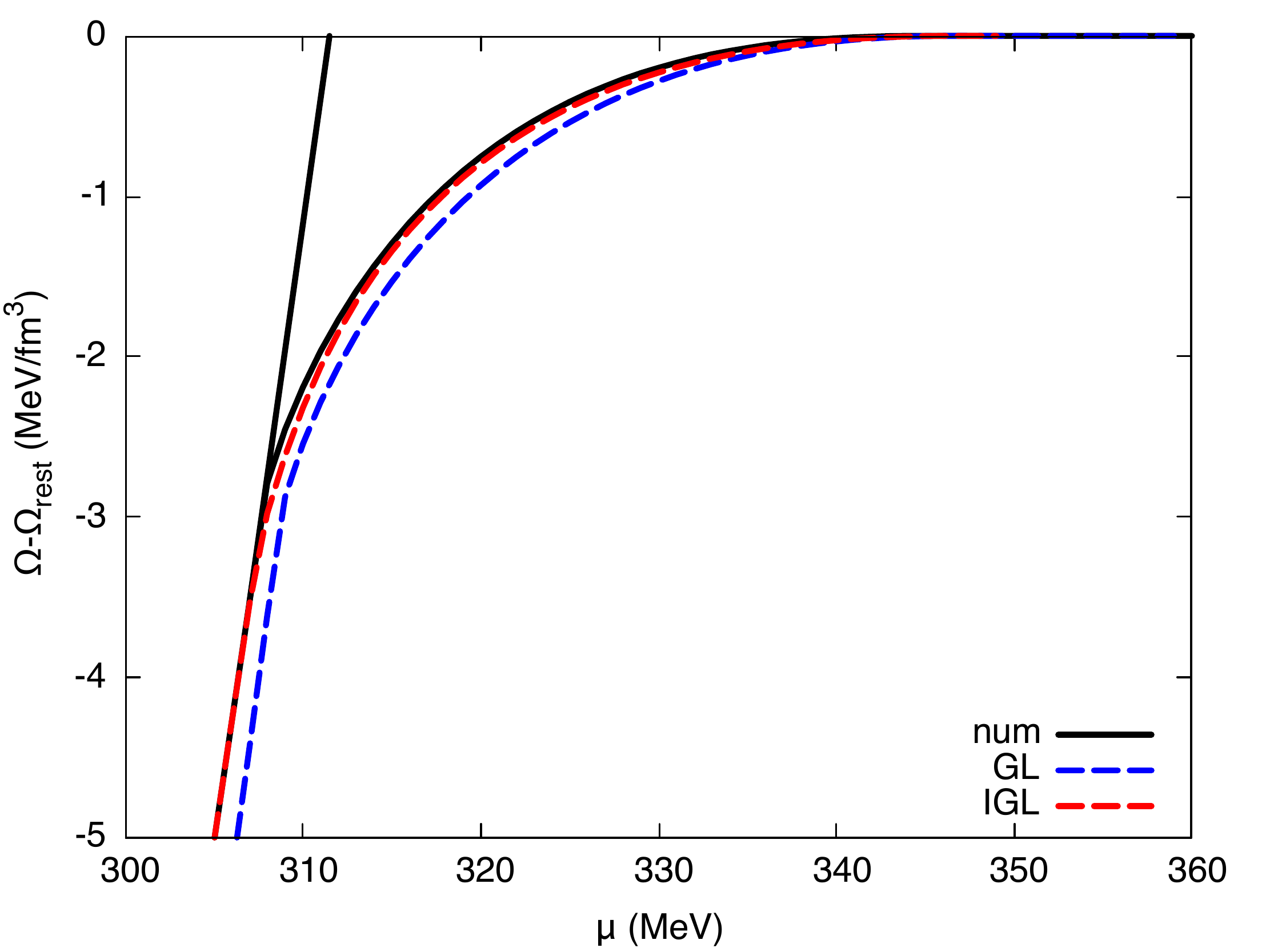}
\caption{Comparison of the numerical results  with  the GL expansion and the IGL approximation  for the RKC, Eq.~\eqref{eq:rkc1d}. Top:  average value of the condensate, $\sqrt{\langle M(z)^2\rangle}$. Bottom: difference  between the free energy at the minimum  and the free energy of the chirally restored phase. \label{fig:TFsolimprov}
}
\end{center}
\end{figure}

After computing the cell averages over $M(z)$ and plugging them in the GL and IGL expression, we obtain the results shown in~\Fig{fig:TFsolimprov}. Again, we find a good agreement of the GL result with the full computation, while the IGL provides a significant quantitative improvement, reproducing the full numerical results within a few percent error.   In this case the effect of the first term in the IGL expansion in Eq.~\eqref{eq:Omega_IGL} is more evident. It forces the average value of the condensate to match the homogeneous value, sensibly improving the agreement with the numerical results. This effect is due to the fact that the RKC ansatz can be seen as a superposition on many different waves with different amplitudes. The long-wavelength amplitudes dominate close to the phase transition with the homogeneous $\chi$SB phase. On the other hand, for a CDW, there is one single spatial frequency $q$, which is large. Therefore in that case the IGL does not improve much with respect to the GL approximation close to this phase transition.

\begin{figure}[h!]
\begin{center}
  \includegraphics[width=.4\textwidth]{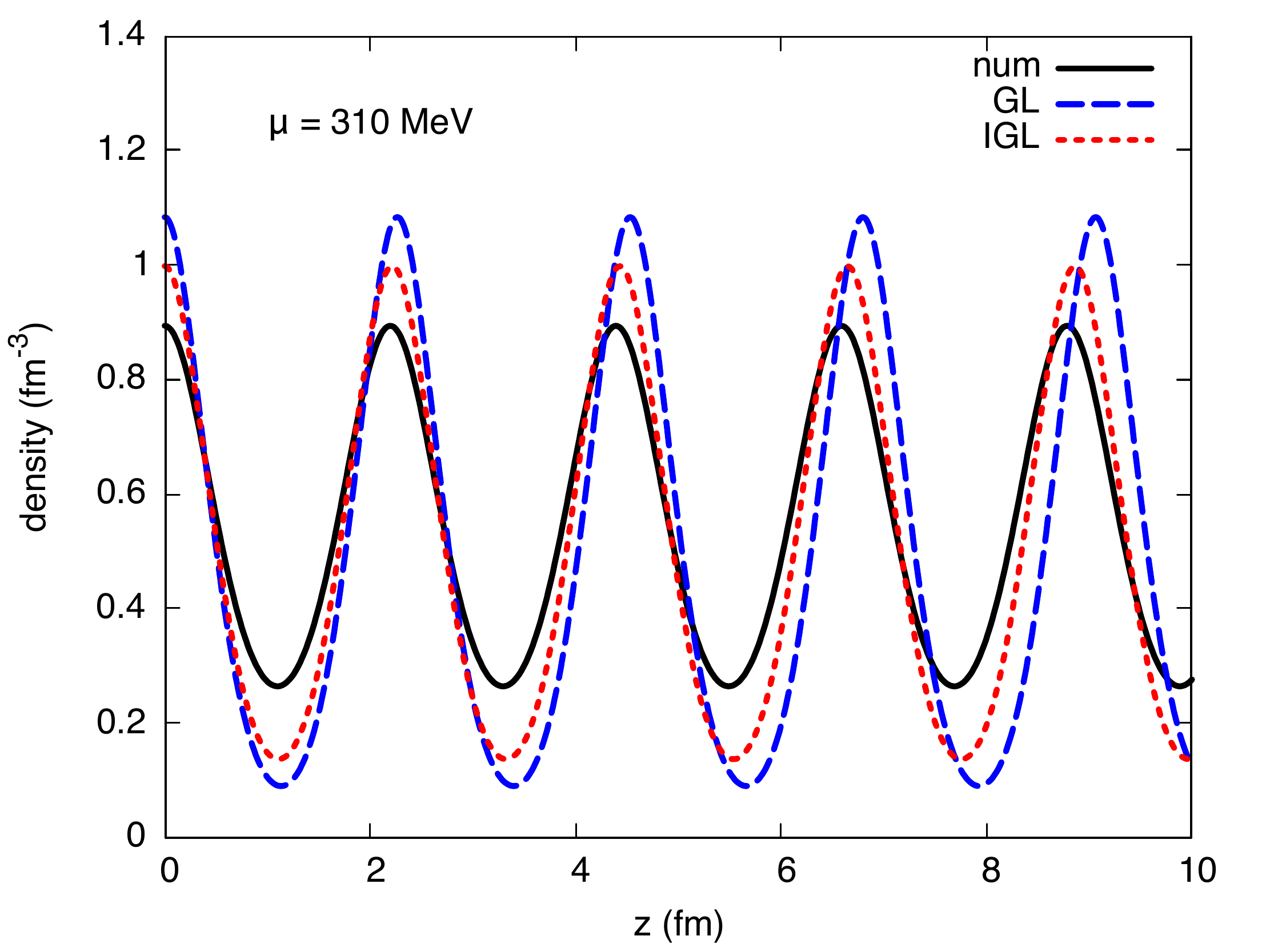}
    \includegraphics[width=.4\textwidth]{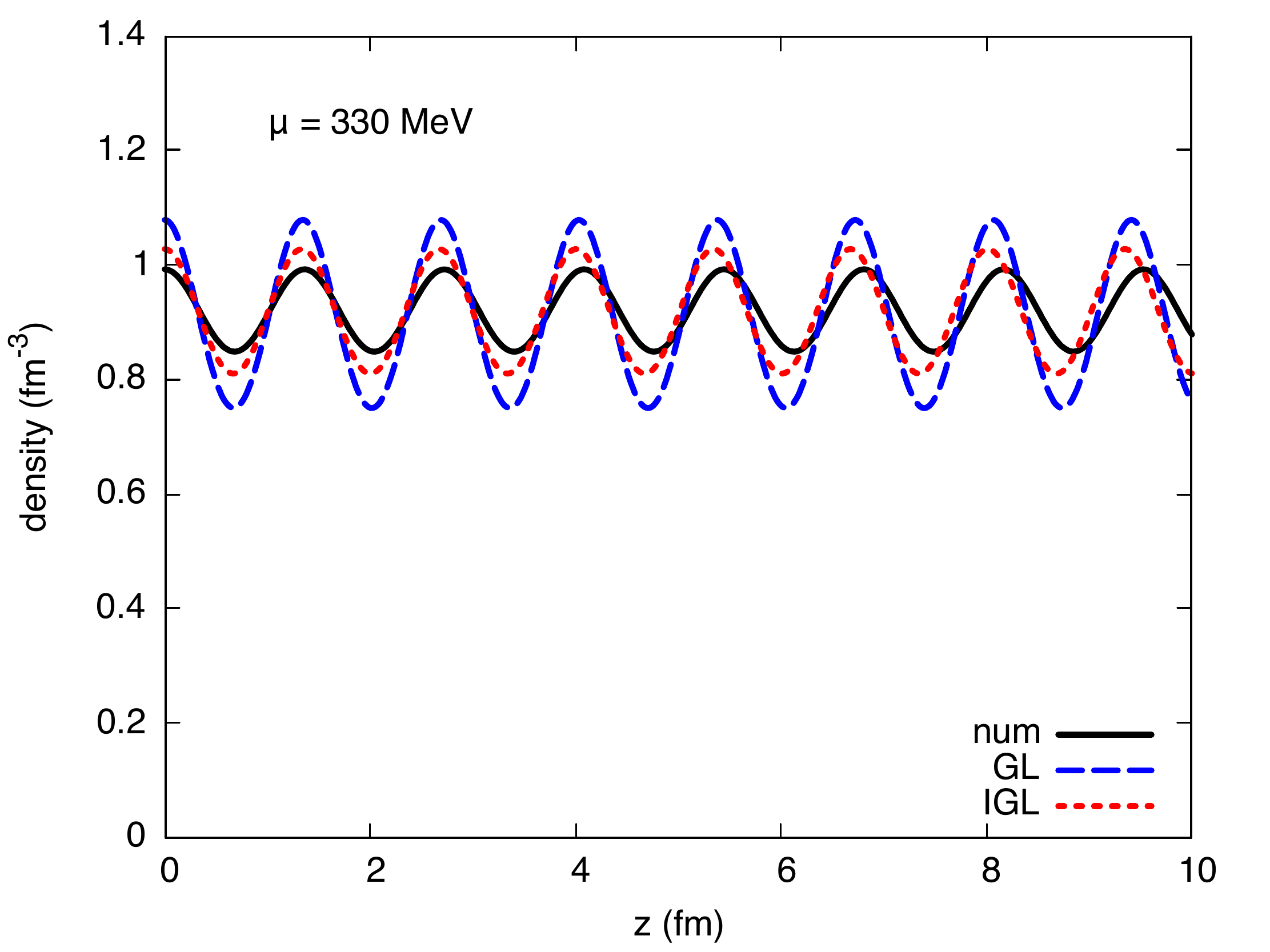}
\caption{Comparison of the quark number density obtained by the GL, IGL and numerical methods   for the RKC modulation \eqref{eq:rkc1d}. 
{The IGL always performs better than the GL, with a discrepancy with the numerical method that is larger close to  the $\chi$SB phase than at the chirally restored phase.}
 The top panel correspond to $\mu=310$ MeV, while the bottom panel is obtained for  $\mu=330$ MeV.
\label{fig:densiRKC}
}
\end{center}
\end{figure}

It might be interesting at this point to compare the spatially modulated quark number density of the system obtained within the GL and IGL approximations with the numerical results of~\cite{CNB:2010}. 
In our case, it is simply obtained by differentiating the integrand of \Eq{eq:Omega_IGL} with respect to $\mu$. 
{This basically amounts to an improvement over a ``local Fermi-gas'' approximation (amounting to simply considering the first term in \Eq{eq:Omega_IGL}), which has already been found to reproduce very well the behavior of the density of the system \cite{Buballa:2015awa}.}
 Our results are shown in \Fig{fig:densiRKC}.  There we can see that once again  the IGL provides a better agreement with the full result as compared to the GL, although in this case the results do not match perfectly, especially close to the  phase transition to the homogeneous broken phase. This is due to the fact that $\Delta$ and $\nu$ give the amplitude and frequency of the density oscillations. Since both are slightly different from the ones obtained by the full numerical calculation, the resulting density has amplitude and oscillation period different from the numerical ones.

The RKC  modulation is somehow similar to the {\it Lasagna phase}  in nuclear matter, that is a type of {\it Pasta phase}~\cite{Ravenhall:1983uh} expected to be realized in the inner crust of compact stars. In this phase nuclei form  sheets immersed in a liquid of nuclear matter. However, with changing densities the Lasagna phase is supposed to be superseded by different {modulations, possibly giving rise to higher-dimensional structures. }  
Following this analogy one {might expect} that higher dimensional modulations can become favored at different values of the quark chemical potential. {Another argument in favor of higher-dimensional modulations comes from quarkyonic matter studies, where it is expected that increasingly complex crystalline structures can be formed by the chiral condensate as the density increases \cite{Kojo:2011}.}

\section{Two-dimensional structures}
\label{sec:twoD}
\begin{figure}
\begin{center}
  \includegraphics[width=.4\textwidth]{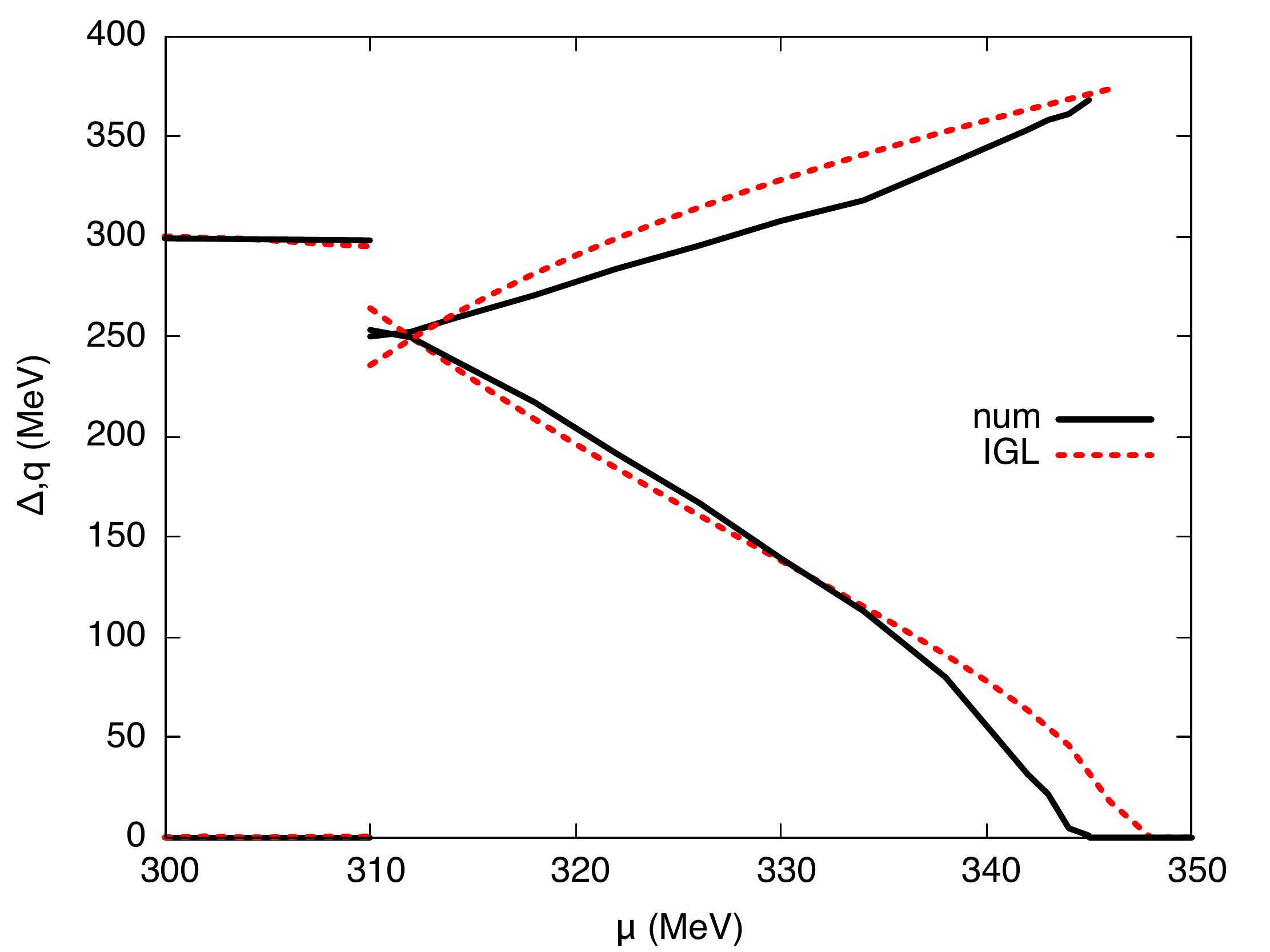}
    \includegraphics[width=.4\textwidth]{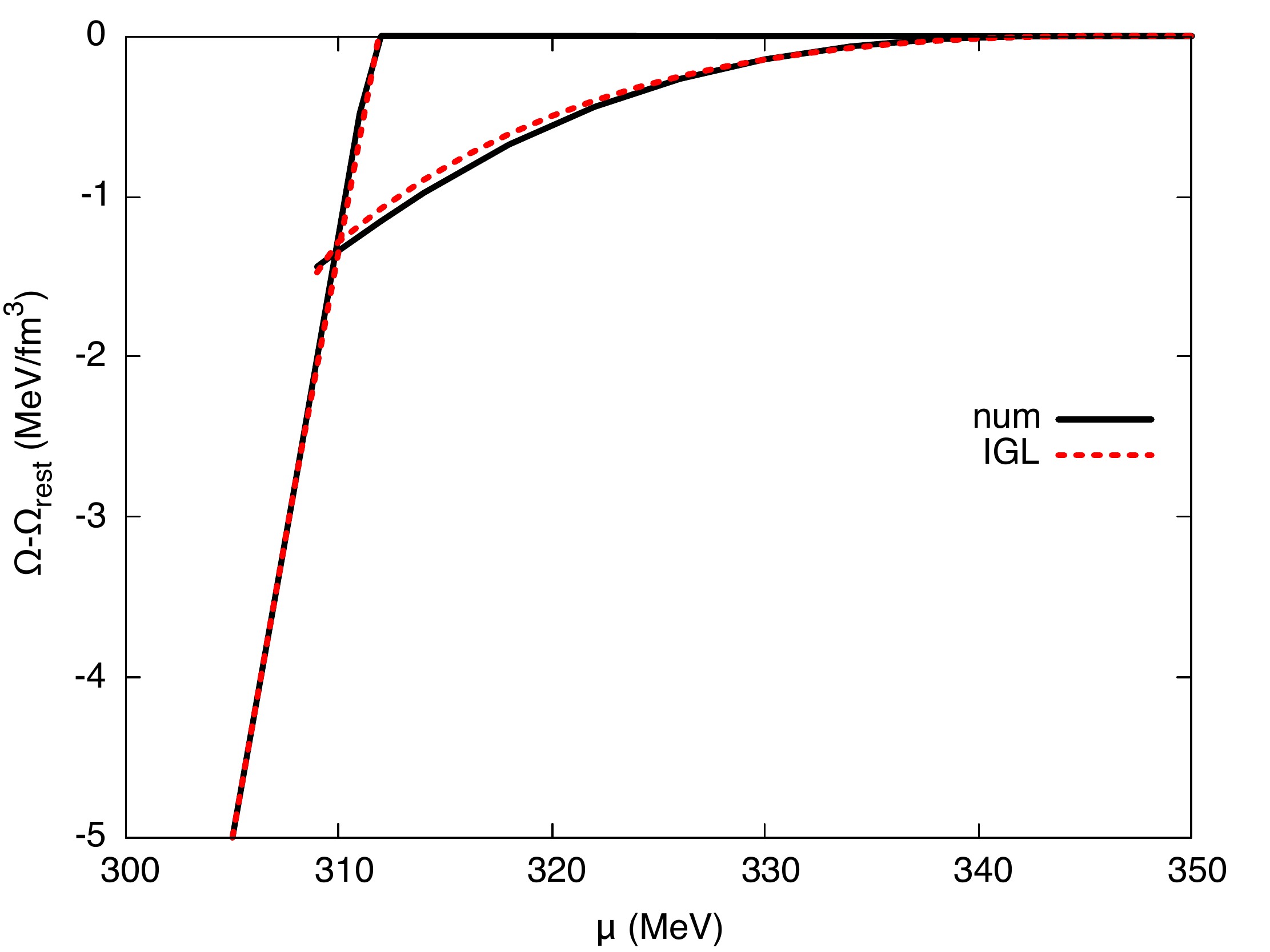}
\caption{Comparison of the  IGL and numerical methods for the  2D cosine modulation ansatz, see  Eq.~\eqref{eq:coscos}, as a function of the quark chemical potential.   Top: Values of the order parameters $\Delta$ and $q$.
 Bottom: comparison of the free-energy difference $\Omega-\Omega_\text{rest}$. 
 \label{fig:cos2d}}
\end{center}
\end{figure}

Since the IGL  provides  
 very accurate results for the order parameters and free energies of one-dimensional modulations with minimal computational effort, let us now move on and consider two-dimensional structures. 
 Close to the Lifshitz point, a systematic GL analysis of different types of higher dimensional modulations has  been performed in~\cite{Abuki:2011pf}, while a complementary numerical analysis for the astrophysical relevant  $T=0$ case can be found in~\cite{Carignano:2012sx}. Comparing the IGL results  with the numerical ones   of~\cite{Carignano:2012sx} for a two-dimensional square lattice with a sinusoidal ansatz, that is, 
 \beq\label{eq:coscos}
 M(x,y) = \Delta \cos(q x) \cos(q y) \,,
 \eeq
we obtain the order parameters and the free energies reported in \Fig{fig:cos2d}.
It is clear that the agreement is again extremely good\footnote{{
 It is worth recalling that the numerical results for the 2D-modulations obtained in \cite{Carignano:2012sx} may carry some numerical uncertainty due to the  cutoffs implemented in the numerical diagonalization of the quark Hamiltonian in momentum space}.}, and we recall that the IGL result can be computed with very limited numerical effort (basically amounting to the evaluation of $\langle\Omega_\text{hom}(\overline{M^2})\rangle${, as all the other terms can be computed analytically}).

Using the IGL method we are in a position to easily test different 2D modulations. First we consider  a square lattice with two RKC-type modulations along the $x$ and $y$ directions, that is 
\beq\label{eq:snsn}
M(x,y) = \Delta\nu {\text{sn}}(\Delta x, \nu)  {\text{sn}}(\Delta y, \nu)\,.
\eeq

\begin{figure}
\begin{center}
  \includegraphics[width=.4\textwidth]{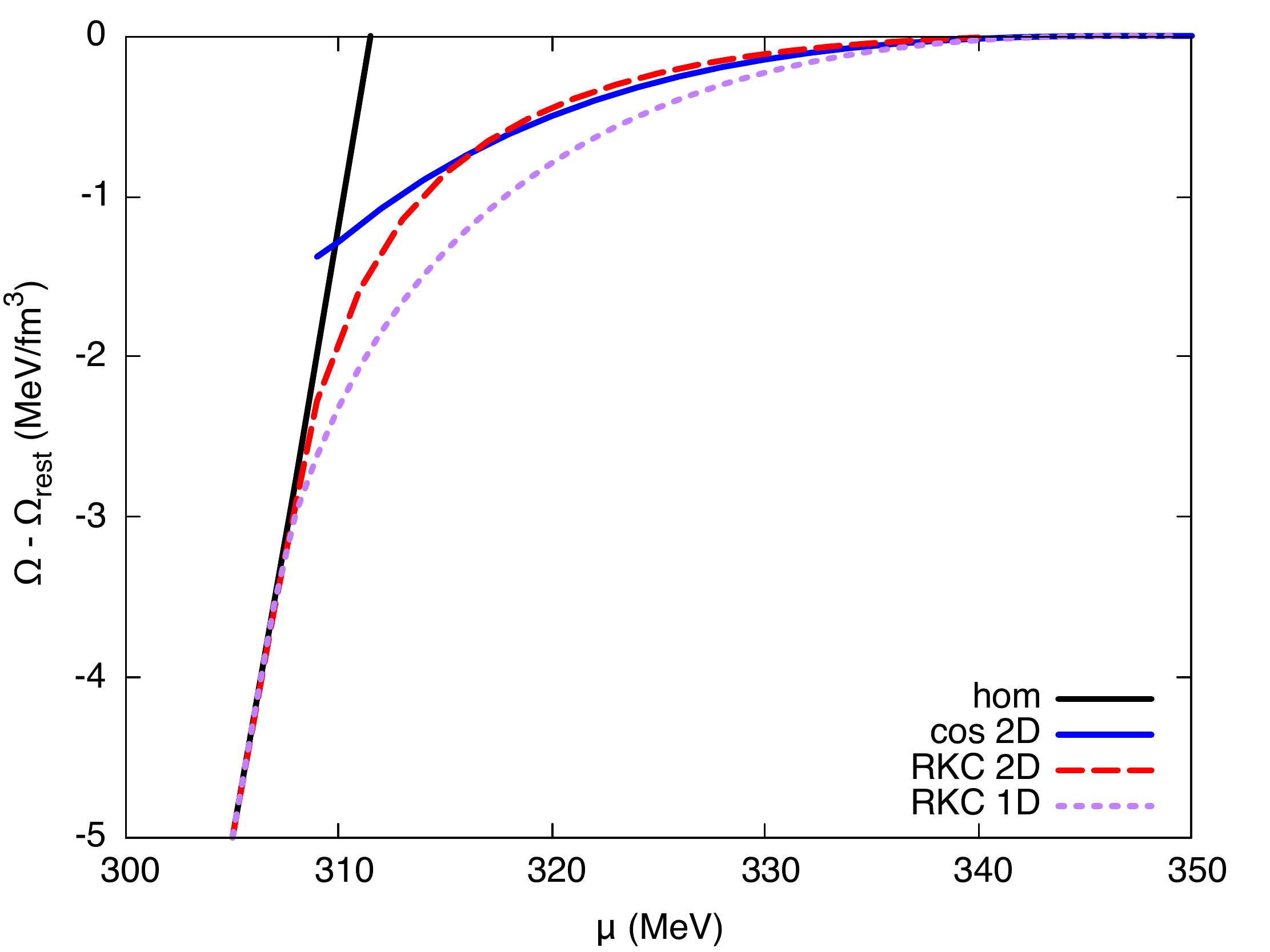}
\caption{Comparison of the free energies for a 1D RKC \eqref{eq:rkc1d},  a 2D cosine \eqref{eq:coscos}  and a 2D RKC \eqref{eq:snsn}, within the IGL approximation. 
 \label{fig:rkc2d}
}
\end{center}
\end{figure}

The practical implementation of this modulation in the numerical framework of~\cite{Carignano:2012sx} would be extremely complicated, as it would in principle require an expansion of the order parameter in a large number of Fourier harmonics and a minimization of the free energy with respect to all of their amplitudes. Instead, within the IGL approximation it can be straightforwardly implemented in the same way as with the 2D cosine. The minimization of the IGL free energy with respect to $\Delta$ and $\nu$ yields qualitatively similar results to the one-dimensional RKC for the order parameters.
When computing the free energy associated with this modulation we find, similarly to what happens with the one-dimensional modulations, that the RKC-type solution is almost degenerate with the cosine one with the exception of the region close to the onset of the inhomogeneous phase, as shown in \Fig{fig:rkc2d}. In that figure we also see that  this type of modulation is also disfavored with respect to its one-dimensional counterpart.  We performed a further check in this direction by considering the ansatz 
\beq
M(x,y) = \Delta \left[ \sqrt{\nu_x}\,  {\text{sn}}(\Delta x, \nu_x)  + \sqrt{\nu_y}\, {\text{sn}}(\Delta y, \nu_y) \right] \,,
\eeq
which can interpolate between a one-dimensional RKC modulation and a more involved two-dimensional structure. Consistent with our other results, we find that the minimum solution always
corresponds to one of the two $\nu$ being zero, while the other reduces to the value obtained when minimizing with the {one-dimensional} ansatz \Eq{eq:rkc1d}.

Thus, as it was already found in \cite{Carignano:2012sx}, we can confirm within our novel approach that 2D modulations are disfavored with respect to 1D modulations at vanishing temperatures. 
{We therefore expect that the same ``hierarchy" found in \cite{Abuki:2011pf} close to the Lifshitz point holds also at vanishing temperatures, and that 3D modulations will thus be even further disfavored compared to two-dimensional ones. } 
 
\section{Qualitative analysis of pairing}
\label{sec:qualitative}
The comparison between  the considered 2D modulations and the 1D  modulations suggests that  the
1D RKC is always favored. This result is in contrast with what is expected to occur in crystalline color superconductors, where a  crystalline 3D pattern seems to be favored~\cite{Alford:2000ze, Mannarelli:2006fy}.
It is believed that in color superconductors the occurrence of the crystalline  phase is due to the maximization of pairing at the Fermi surface, indeed the presence of a collective Fermi surface phenomenon seems to be the key-point for obtaining a crystalline phase.

\begin{figure*}[t!]
\centering
\includegraphics[width=.3\textwidth]{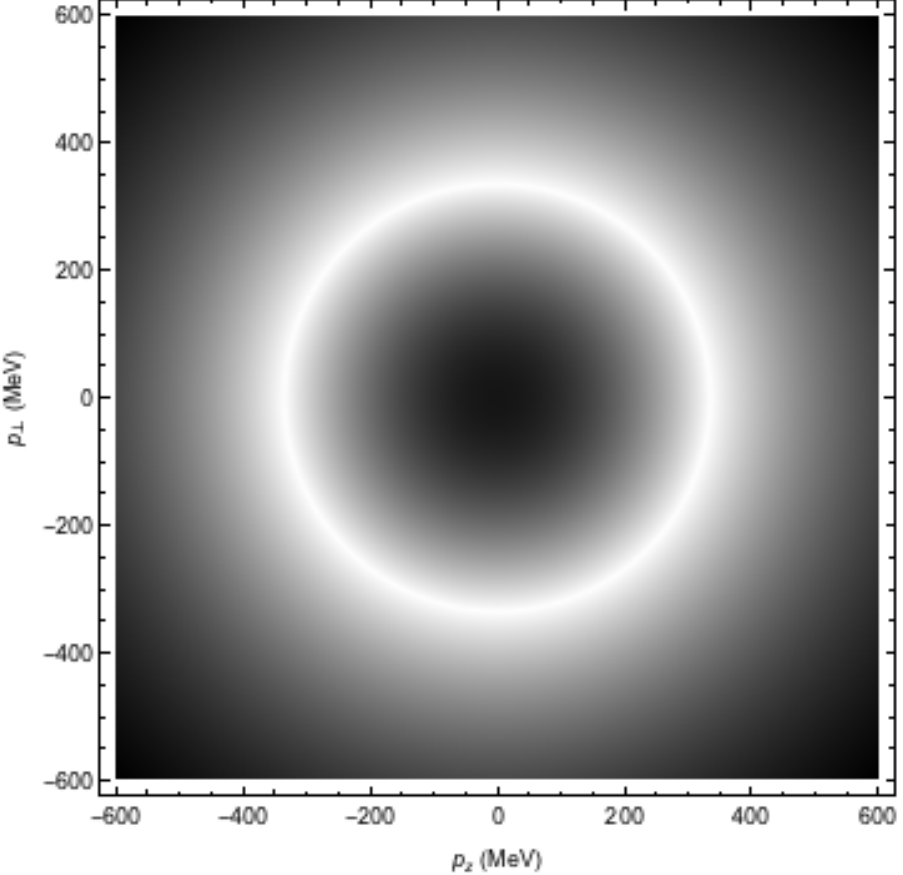}\,\,\,\,\,\,;
\includegraphics[width=.3\textwidth]{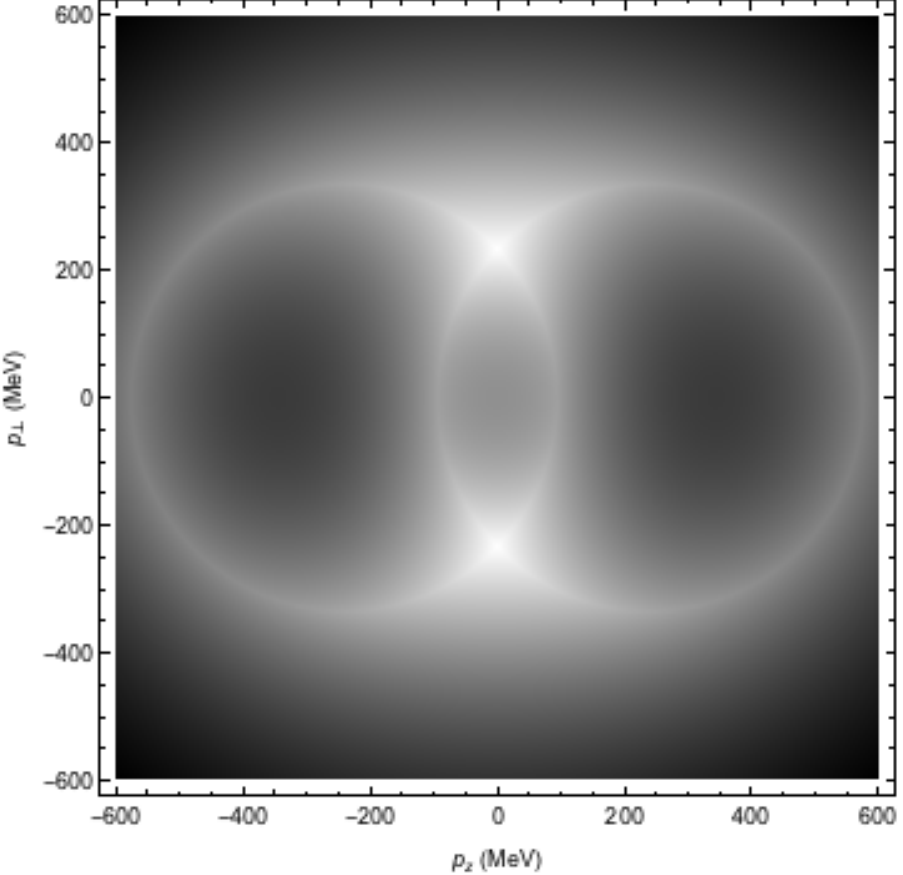}\,\,\,\,\,\,
\includegraphics[width=.3\textwidth]{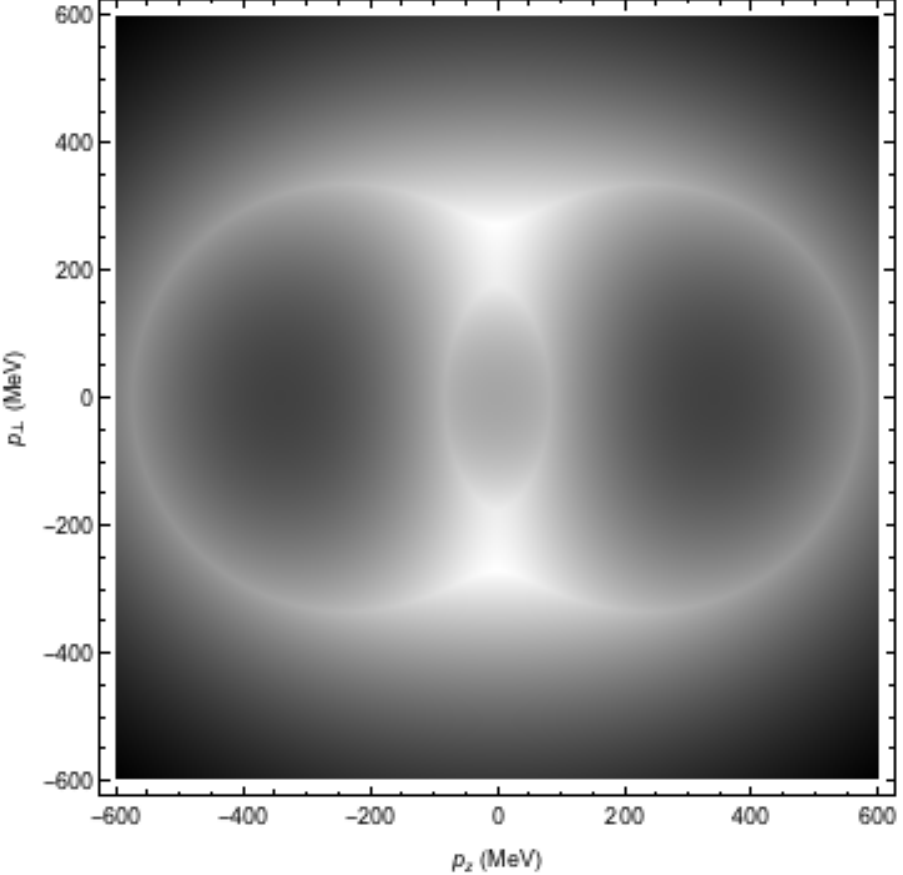}
\caption{Two dimensional contour plots of the integrand of the free energy for CDW ansatz, \Eq{eq:Omega_CDW}. All the results are obtained for  $\mu=335$ MeV, but different values of $q$ and $\Delta$. The lighter region corresponds to the  region where the free energy cost for exciting quasiparticles is smaller.  Left:  unpaired phase, $q=0$ and $\Delta=0$.  Center:  $q= 241$ MeV  and $ \Delta=0$. The effect of the large momentum $q$ is to strongly displace the Fermi spheres.  Right:   $q= 241$ MeV  and  $\Delta=44$ MeV (corresponding to the energetically favored values at $\mu=335$ MeV). The smearing of the lighter region is  due to the pairing.
  }\label{fig:integrand}
\end{figure*}
Quite generally, a certain modulation is energetically favored if  the  energy gain due to pairing is larger than the energy cost of having pairs with nonvanishing total momentum. Let us examine in detail what happens in an inhomogeneous $\chi$SB condensate. For a qualitative understanding of the phenomenon we consider first the effect of a nonvanishing momentum and then we allow for pairing. For understanding whether multidimensional pairing is favored, we consider what happens for a plane wave ansatz. 
{As discussed in~\cite{Mannarelli:2006fy,Anglani:2013gfu}, one way of representing the Fermi surface effects is to inspect the integrand of  the free energy, corresponding, in our case, to the integrand appearing in \Eq{eq:Omega_CDW}. }
 In  Fig.~\ref{fig:integrand} we plot this function  at $\mu=335$ MeV, that is within the inhomogeneous $\chi$SB window. The left panel corresponds to the  free case, that is, $q=0$ and $\Delta=0$. 
 The integrand is peaked at $p=\mu$, meaning that the larger contribution comes from the Fermi surface, corresponding to the lighter region in Fig.~\ref{fig:integrand}. This is  the so-called {\it pairing region}, while the parts well inside the Fermi sphere or well outside it correspond to the {\it blocking regions} (see the discussion in \cite{Mannarelli:2006fy} about pairing and blocking regions in color superconductors). In other words, pairing  well inside/outside the Fermi sphere has a large free-energy cost, because particles should climb to the tip of the free energy (integrand), which is at the Fermi sphere. On the other hand, particle and hole excitations at the Fermi sphere are already at the tip of the mountain, that is  to the largest possible energy, and they can eventually pair at no cost to form a chiral condensate~\cite{Kojo:2009}.

Now we consider  a  momentum shift of the fermions. When pairs have nonvanishing total momentum, one can imagine first to displace fermions by $q$ and then to turn  on pairing. This is exactly what  one does  when diagonalizes the full Hamiltonian for the single plane wave to obtain the free energy in \Eq{eq:Omega_CDW}.
This procedure is discussed in detail in~\cite{Mannarelli:2006fy} for crystalline  color superconductors, where it is shown how by a proper momentum shift
 the quark propagator becomes diagonal (see also \cite{Buballa:2014tba} for an analogous discussion for inhomogeneous chiral condensates). 
This momentum shift has the effect of separating the Fermi spheres, as shown in the central panel of \Fig{fig:integrand}  for $q=241$ MeV. Now, the only pairing region corresponds to the ribbon where the two Fermi spheres touch. This picture also explains why $q<\mu$, indeed if this were not the case the two Fermi spheres would have no overlapping regions. Exciting quasiparticles and/or holes in the pairing ribbon has no free energy cost, whereas particles from all other regions should climb an energy barrier. Indeed, we see that this is exactly what happens for $\Delta=44$ MeV, right panel, where the smearing of the touching regions is exactly due to pairing. It is not possible now to  add pairing in different regions of the Fermi spheres, say in the region $p_\perp \sim 0$, because these regions  are too far apart and therefore the free energy cost for exciting particles and/or holes would be too high. Therefore, multidimensional modulations are disfavored in the $\chi$SB phase because $q$ is too large. 

\begin{figure}
\includegraphics[width=.4\textwidth]{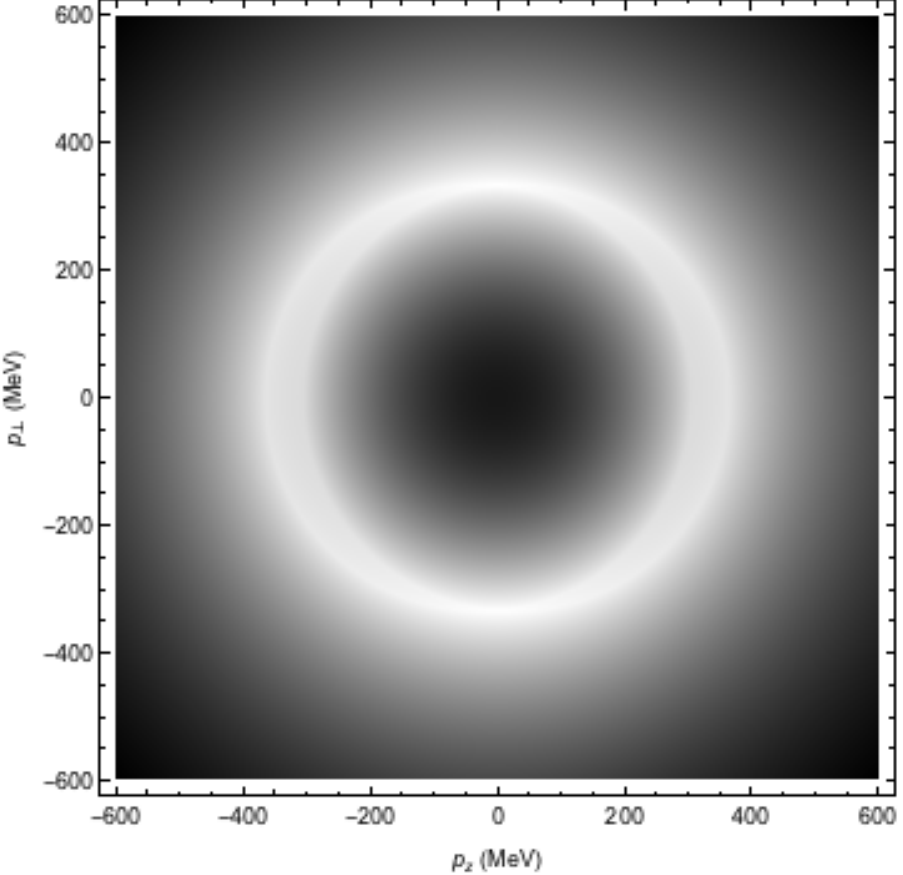}
\caption{Contour plots of the integrand of the free energy for CDW ansatz, Eq.~\eqref{eq:Omega_CDW}. For  $\mu=335$ MeV, $q= 30$ MeV  and $ \Delta=5$  MeV. }\label{fig:integrand_2}
\end{figure}
This does not happen in color superconductors. Indeed, one important difference between the inhomogeneous $\chi$SB phase and the crystalline color superconductors, regards the magnitude of  $q$. Broadly speaking, $q$ has to be proportional to the stress exerted on the pairing mechanism. In the $\chi$SB phase the stress is proportional to $\mu$, because pairing is related to the formation of a chiral condensate. 
On the other hand, in color superconductors $q \propto \delta\mu$, where $\delta\mu\ll\mu$ is the mismatch between the Fermi spheres due to an unbalance between quarks of different flavors.
For illustrative purposes, let us consider the non-energetically-favored $\chi$SB configuration corresponding to small $q$ and $\Delta$,  somehow mimicking what happens in color superconductors. We show in Fig.~\ref{fig:integrand_2}  the integrand of Eq.~\eqref{eq:Omega_CDW},  with $q=30$ MeV and $\Delta=5$ MeV. Again, pairing can happen in the ribbon where the two Fermi spheres overlap. However, it is clear now that it would be possible to slightly modify the Fermi sphere for allowing pairing, say in the   $p_\perp \sim 0$ plane, at a small free-energy cost.

\section{Conclusions}
\label{sec:conclusions}
We have presented a novel approach to {study spatially} inhomogeneous pairing by an improved Ginzburg-Landau expansion,  Eq.~\eqref{eq:Omega_IGL}.  
This approach relies on a scale separation between long-wavelength fluctuations, dominating the transition to the homogeneous phase, and rapid fluctuations governing the transition to the chirally restored phase. 
{The IGL reproduces correctly by construction the homogeneous limit and allows for a description of the chiral restoration transition from the inhomogeneous phase with
arbitrarily high precision by a controlled gradient expansion.}

We have applied the IGL to the study of the inhomogeneous $\chi$SB
 phase at $T=0$, reproducing the results obtained by numerical methods and extending the analysis to novel  structures. 
  These structures can hardly be studied  by the numerical method, because of the complicated Fourier expansion technique underlying these methods. On the other hand, the IGL expansion turns out to be an extremely powerful tool, allowing us to quickly examine various crystalline structures to an arbitrarily accurate approximation. In this way we checked that  various 2D modulations are disfavored
with respect to the 1D RKC one in Eq.~\eqref{eq:rkc1d}, confirming and extending previous results obtained via brute-force numerical computations \cite{Carignano:2012sx}. 

 It is worth emphasizing that  no approximate method so far has been used to analyze the $T=0$ case, probably because the standard GL approximation was assumed to be unreliable. Actually, we find that the GL approximation at the ${\cal O} (\alpha_8)$ gives {a surprisingly good qualitative agreement with the results of the full numerical computations}. However, to {make a quantitative comparison of} the free energies of different structures a refined approach must be used, and the IGL devised here performs this task {excellently}. 
 In particular, {we showed that it is able to give} an accurate description of both the second order phase transition to the chirally restored phase and of the phase transition to the homogeneous $\chi$SB phase.  {As it turns out, a  small number of additional specific gradient terms is enough to provide an excellent agreement with the numerical data. }

{Finally, it is worth recalling that fluctuations are expected to have a strong effect on the formation of inhomogeneous condensates~\cite{Lee:2015bva, Yoshiike:2017kbx, Hidaka:2015xza}, particularly in the case of lower-dimensional modulations \cite{Landau:1969} (see also \cite{Buballa:2014tba} for a discussion). The inclusion of fluctuations in the  IGL framework would lead to a systematic improvement beyond the mean-field approximation.}

The present work can be extended in many different ways. The IGL free-energy can be
used to rapidly evaluate the free energy of various crystalline color superconducting configurations, as the ones considered in \cite{Rajagopal:2006ig}, and to extend the analysis to novel modulations. The only modifications needed in  Eq.~\eqref{eq:Omega_IGL} are the replacement of the free energy of the homogeneous phase with the 2SC one (for two flavors) or of the color flavor locked one (for three flavors)  and to replace the $\alpha_n$ coefficients with the pertinent ones{, which can in principle be obtained by considering  modulations for which the eigenvalue spectrum is known, such as a simple Fulde-Ferrell  type plane wave~\cite{FF,LO,Anglani:2013gfu}. In this case one can also compare the IGL results with those obtained by the numerical method in~\cite{NB:2009}. We will shortly present results on  this topic. 

Moreover, the IGL can be modified to  simultaneously include the chiral and diquark condensates for examining the coexistence of the  inhomogeneous $\chi$SB and of the crystalline color superconducting phase. In this case, the color-superconducting phase is expected to arise where the chiral condensate is small, or equivalently,  where the density is large. Since  1D chiral modulations are favored, we expect that a cosine  modulation, see for example~\cite{Anglani:2013gfu}, could be favored.

%

\end{document}